\documentclass[reprint,amsmath,amssymb,aps,prb,showpacs,floatfix,superscriptaddress]{revtex4-1}
\usepackage{graphicx}
\usepackage{bm}
\usepackage{multirow}
\usepackage{verbatim}
\usepackage{longtable}
\usepackage{color}
\usepackage{hyperref}
\usepackage[normalem]{ulem}

\newcommand{\e}{e}
\newcommand{\SSS}{\mathbf{S}}
\newcommand{\VVV}{\mathbf{V}}
\newcommand{\JJJ}{\mathbf{J}}
\newcommand{\WWW}{\mathbf{W}}
\newcommand{\GGG}{\mathbf{G}}
\newcommand{\TTT}{\mathbf{T}}
\renewcommand{\i}{i}
\renewcommand{\figurename}{Fig.}

\newcommand{\sectionname}{Section}
\newcommand\abs[1]{\lvert#1\rvert}
\newcommand\absB[1]{\left\lvert#1\right\rvert}

\newcommand\bra[1]{\langle#1\rvert}
\newcommand\ket[1]{\lvert#1\rangle}

\newcommand{\Tr}{\mathrm{Tr}}

\newcommand{\ve}{\varepsilon}

\newcommand{\pd}{\partial}
\newcommand{\m}{\mathcal}
\newcommand{\upbl}{{\color{black}\uparrow}}
\newcommand{\downbl}{{\color{black}\downarrow}}
\newcommand{\etabl}{{\color{black}s}}
\newcommand{\up}{\uparrow}
\newcommand{\down}{\downarrow}

\newcommand{\bk}{{k}}

\newcommand{\bq}{{q}}
\newcommand{\bkap}{\bq}

\newcommand{\cd}{c^{\dag}}
\newcommand{\can}{c^{\phantom{\dag}}}

\newcommand{\dd}{d^{\dag}}
\newcommand{\dan}{d^{\phantom{\dag}}}
\newcommand{\gamd}{\gamma^{\dag}}
\newcommand{\gaman}{\gamma^{\phantom{\dag}}}

\newcommand{\LR}{\mathrm{LR}}

\newcommand{\D}{\mathrm{D}}
\newcommand{\T}{\mathrm{T}}

\newcommand{\mrN}{\mathrm{N}}
\newcommand{\mrce}{\mathrm{ce}}
\newcommand{\mri}{\mathrm{d}}
\newcommand{\s}{\sigma}
\newcommand{\ia}{\alpha\bk}
\newcommand{\iap}{\alpha'\bk'}
\newcommand{\absDelta}{\Delta}

\newcommand\sgn{\mathrm{sgn}}

\begin{document}

\title{Yu-Shiba-Rusinov states in phase-biased S-QD-S junctions}

\author{Gediminas Kir\v{s}anskas}
\affiliation{Center for Quantum Devices, Niels Bohr Institute, University of Copenhagen, DK-2100 Copenhagen \O, Denmark}

\affiliation{Mathematical Physics and NanoLund, Lund University, Box 118, 221 00 Lund, Sweden}

\author{Moshe Goldstein}
\affiliation{Raymond and Beverly Sackler School of Physics and Astronomy, Tel Aviv University, Tel Aviv 69978, Israel}

\author{Karsten Flensberg}
\affiliation{Center for Quantum Devices, Niels Bohr Institute, University of Copenhagen, DK-2100 Copenhagen \O, Denmark}

\author{Leonid I. Glazman}
\affiliation{Department of Physics, Yale University, New Haven, CT 06520, USA}

\author{Jens Paaske}
\affiliation{Center for Quantum Devices, Niels Bohr Institute, University of Copenhagen, DK-2100 Copenhagen \O, Denmark}

\date{\today}
\begin{abstract}
We study the effects of a phase difference on Yu-Shiba-Rusinov (YSR) states in a spinful Coulomb-blockaded quantum dot contacted by a superconducting loop. In the limit where charging energy is larger than the superconducting gap, we determine the subgap excitation spectrum, the corresponding supercurrent, and the differential conductance as measured by a normal-metal tunnel probe. In absence of a phase difference only one linear combination of the superconductor lead electrons couples to the spin, which gives a single YSR state. With finite phase difference, however, it is effectively a two-channel scattering problem and therefore an additional state emerges from the gap edge. The energy of the phase-dependent YSR states depend on the gate voltage and one state can cross zero energy twice inside the valley with odd occupancy. These crossings are shifted by the phase difference towards the charge degeneracy points, corresponding to larger exchange couplings. Moreover, the zero-energy crossings give rise to resonant peaks in the differential conductance with magnitude $4e^2/h$. Finally, we demonstrate that the quantum fluctuations of the dot spin do not alter qualitatively any of the results.
\end{abstract}

\maketitle

\section{\label{sec:intro}Introduction}

Yu-Shiba-Rusinov~\cite{Yu1965, Shiba1968, Rusinovfirst} states are bound states in an $s$-wave superconductor induced inside the energy gap by local magnetic moments. Individual localized Yu-Shiba-Rusinov (YSR) states have been observed both by scanning tunneling spectroscopy of magnetic atoms like Mn or Cr adsorbed on superconducting Pb or Nb substrates,~\cite{Yazdani1997, Ji2008, Ji2010, Franke2011, Balatsky2006} and as subgap states in Coulomb blockaded quantum dots (QD) coupled to superconducting (S) leads.~\cite{Eichler2007, Grove-Rasmussen2009, Deacon2010, Pillet2010, Chang2012, Lee2012, Kim2013, Kumar2014,Lee2014} The quantum dot realization is based on the spin-$1/2$ of odd-occupation charge states, and is therefore free of most of the material dependent complications for adatoms on a surface, like mixed valence, higher spin, and magnetic anisotropy. Furthermore, the quantum dot system allows for electrical tunability of the particle-hole asymmetry and, to some extent, the exchange coupling between the spin on the quantum dot and the quasiparticles in the superconductor, which makes it an ideal system for studying the properties of individual YSR states.
\begin{figure}[!ht]
\includegraphics[width=0.85\columnwidth]{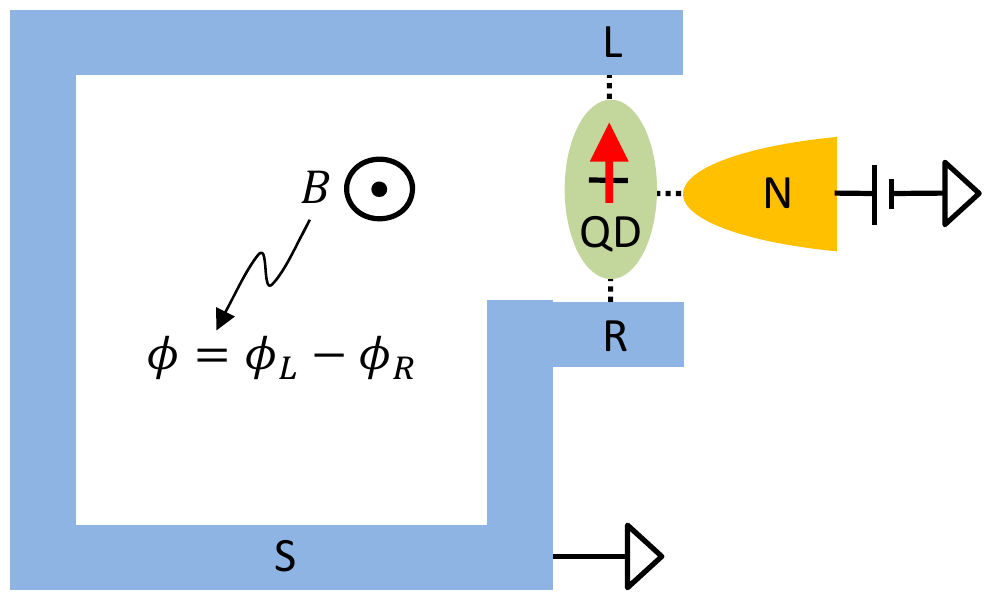}
\caption{\label{fig1}(Color online) Sketch of the device studied here  (following the experiments in Refs.~\onlinecite{Pillet2010, Chang2012}), comprised of a Coulomb-blockaded QD (green), tunnel coupled to superconducting leads (blue) with an applied phase difference, $\phi$, controlled by a magnetic flux. A backgate voltage is adjusted to provide a single spin-1/2 on the QD, and the resulting phase-dependent YSR states are probed by a normal-metal tunnel probe (yellow).}
\end{figure}

Recently, sharp subgap states have been observed with a weakly coupled normal (N) metal lead probing an S-QD-S junction as illustrated in \figurename~\ref{fig1}, where the quantum dot was formed in a Coulomb-blockaded segment of carbon nanotube~\cite{Pillet2010} or InAs nanowire~\cite{Chang2012} spanning a superconducting Al loop. The gate control allows for determination of even (spinless) or odd (spinful) charge states of the quantum dot by even-odd effects (such as absence or presence of Kondo resonances~\cite{Chang2012}) of the stability diagram. By tuning the magnetic flux piercing the Al loop, one may control the phase-difference across the quantum dot, and this device therefore provides additional information about the phase dependence of the subgap states. It is this phase dependence of the spin-induced YSR states, which is addressed in the present paper.

Earlier measurements of supercurrent through phase-biased Coulomb-blockaded quantum dots,~\cite{Cleuziou2006, VanDam2006, Joergensen2007, Eichler2009} have already demonstrated that odd-occupied spin-degenerate dots may lead to negative ($\pi$-phase) supercurrent. Also, a supercurrent sign reversal, i.e., a $\pi-0$ transition, has been shown to take place when adjusting the gate voltage to move away from odd occupancy, thereby increasing the ratio of the Kondo temperature $T_K$ to the superconducting gap $\Delta$, in accordance with a number of theoretical predictions.~\cite{Kulikfirst, Bulaevskiifirst, Bulaevskii1978, Glazmanfirst, Spivak1991, Rozhkov1999, Siano2004, Choi2004, Karrasch2008} Whereas the supercurrent only carries information about the ground state, the additional spectroscopic information from experiments like Refs.~\onlinecite{Pillet2010, Chang2012} now offers an opportunity to learn more about the subgap \textit{excitations} caused by a local magnetic moment.

The nature of subgap states depends on the ratio of the charging energy $E_C$ and the superconducting gap $\Delta$. For $E_C>\Delta$ and $E_C\gg\Gamma$, where $\Gamma$ denotes the elastic broadening of the dot states due to their coupling to the leads, the natural starting point for the spinful dot is the Kondo model \cite{Pustilnikreview} with exchange coupling $J$ and with the lead conduction electrons described by BCS Hamiltonians. As we show below, in absence of a phase difference across the junction, this reduces to a single-channel YSR problem. For weak exchange coupling, $T_{K}\ll\Delta$, where the ground state is a doublet, the YSR state is an excited singlet state consisting of a single quasiparticle in the lead bound to the dot spin. As the exchange coupling is increased the excited state crosses zero energy and the ground state changes abruptly to spin singlet~\cite{Soda1967, Satori1992, Bauer2007} at $T_{K}\sim \Delta$, where $T_K$ is the Kondo temperature. For even larger values $T_K\gg\Delta$, the ground state can be described as the well-known Kondo singlet.

\begin{figure}
\includegraphics[width=0.8\columnwidth]{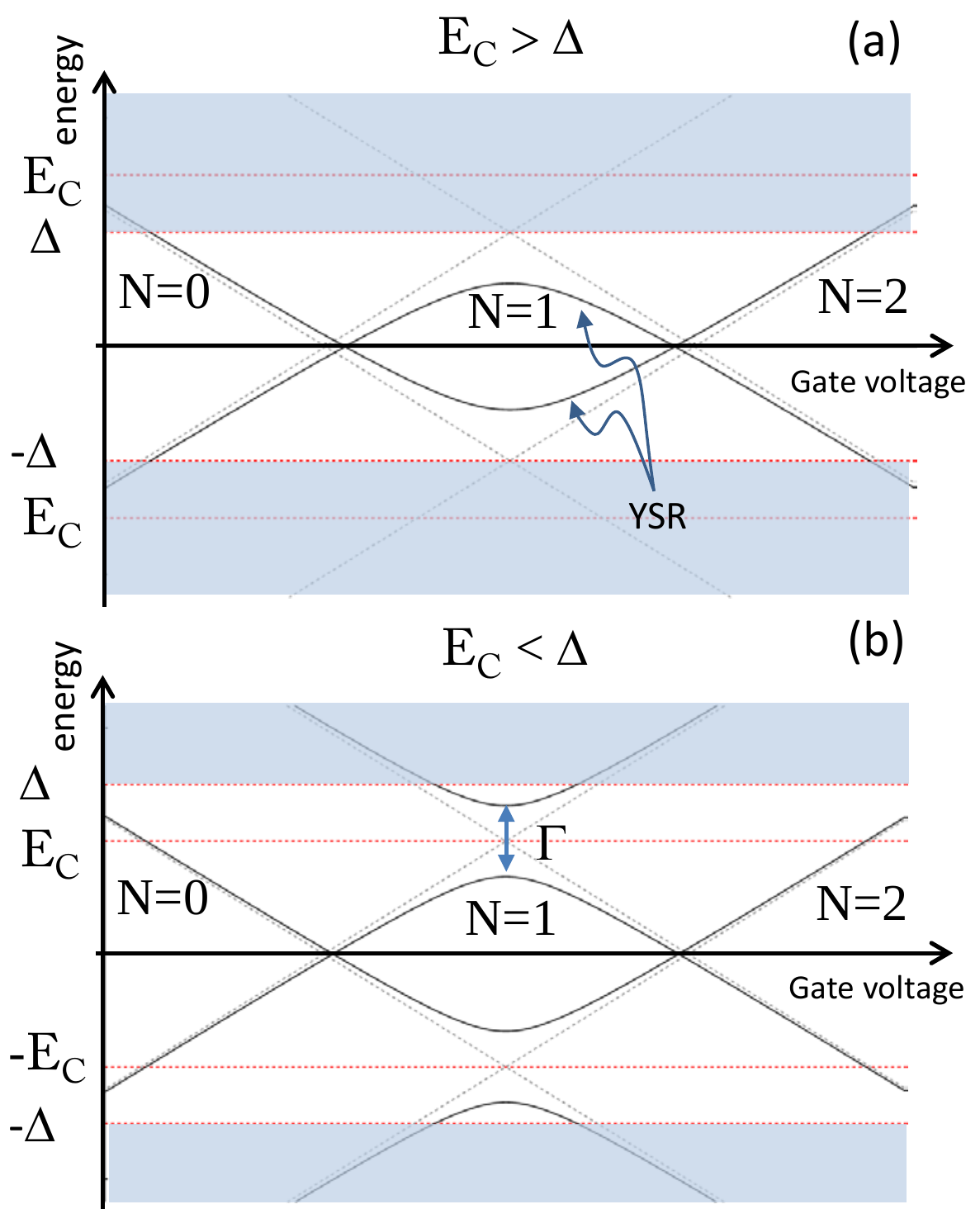}\\
  \caption{\label{fig2} (Color online) Schematics of the two situations, $E_C>\Delta$ and $E_C<\Delta$. In both cases there is a subgap state in the odd diamond. For large $\Delta$ it appears because of hybridization of the $N=0$ and $N=2$ charge states by the superconductor, while for large $E_C$ it appears because of hybridization between the dot electron and quasiparticles in the leads, forming a bound state, which is the YSR state. In both cases the average occupation on the dot is equal to one and the subgap states have similar dependences on gate voltage. In this paper, we focus on the situation (a) with $E_C>\Delta$.}
\end{figure}

In the opposite limit, $E_C<\Delta$, the natural starting point for understanding the subgap states is a model where the superconducting electrons are integrated out, giving rise to a local pairing on the dot with strength $\Gamma$, which was studied by Meng \textit{et al.}~\cite{Meng2009} It gives a mixing of states with occupation $N\pm 1$ when the average occupation is $N$. With odd average occupation there will be two subgap states split by $\Gamma$, as found already in the single orbital Anderson model.~\cite{Meng2009} Even though the two cases $E_C<\Delta$ and $E_C>\Delta$ are naturally described in different languages, the physical situations are similar. In both cases, the excited states correspond to an extra quasiparticle bound by the local spin. For $E_C>\Delta$ the bound particle resides mainly in the superconductor in the form of a YSR state, while for $E_C<\Delta$ it resides mainly on the dot because of the hybridization of 0 and 2 electrons.
The two situation are illustrated in \figurename~\ref{fig2}. For the experiments reported in Refs.~\onlinecite{Chang2012} and \onlinecite{Lee2014} the relevant limit is $\Delta<E_C$, which is also the limit considered in this paper.

For a multi-level quantum dot, the finite level spacing, $\delta E$, will also affect the simple evolution of ground states when $\delta E\sim T_{K},\Delta$.
In this paper we thus restrict our attention to small dots, for which the level spacing is the largest energy scale, and $\Gamma,\Delta\ll E_{C}\sim\delta E$, complementary to the $\Delta\gg E_{C}$ regime studied by Meng \textit{et al.}~\cite{Meng2009} Furthermore, we restrict our attention to gate voltages adjusted to accommodate odd occupation and hence a net spin-1/2 on the quantum dot, which will be described within an effective cotunneling (Kondo) model. For the main part of the paper, however, the dot spin is treated as classical (non-fluctuating), being polarized in a fixed direction. We will show that this approximation does not modify the physics substantially (at least for weak dot-lead coupling), while considerably simplifying the calculations.

The paper is organized as follows. In \sectionname~\ref{sec:model} an effective Kondo model for an odd-occupied quantum dot coupled to superconducting leads is introduced. In \sectionname~\ref{sec:polar}, we derive the subgap states within the polarized-spin approximation. The corresponding nonlinear tunneling conductance to an auxiliary normal lead is discussed in \sectionname~\ref{sec:condnmp}, with a few technical details relegated to \appendixname~\ref{sec:appendixAB}, and finally the supercurrent through the phase-biased S-QD-S junction is addressed in \sectionname~\ref{sec:scrt}. In \sectionname~\ref{sec:Q}, we briefly compare the results from the polarized-spin approximation with the perturbative (in dot-lead coupling) results including the full quantum dynamics of the dot spin, which we derive in \appendixname~\ref{sec:sbgstQS}.

\section{\label{sec:model}The Model}

We consider a quantum dot tunnel-coupled to two superconducting leads, capacitively coupled to a gate electrode, and subjected to an applied magnetic field. The coupling to the normal metal electrode is also included for the purpose of doing transport spectroscopy.

The quantum dot connected to these three leads is modelled by a single orbital Anderson-type model:
\begin{equation}
H=H_{\mrN}+H_{\LR}+H_{\D}+H_{\T}.
\end{equation}
The superconducting leads are described by the effective Bardeen-Cooper-Schrieffer (BCS) mean-field Hamiltonian
\begin{align}\label{hamLR}
    H_{\LR}=&\sum_{\ia\s}\xi_{\bk\s}^{{}}\cd_{\ia\s}\can_{\ia\s}\\
    &-\sum_{\ia}\left(\Delta_{\alpha}\cd_{\ia\up}\cd_{\alpha,-\bk\down}
    +\Delta_{\alpha}^{*}\can_{\alpha,-\bk\down}\can_{\ia\up}\right),\nonumber
\end{align}
where $\xi_{\bk\s}^{{}}=\xi_{\bk}+\s g_{\mrce}B/2$ and $\xi_{\bk}^{{}}=\ve_{\alpha\bk}-\mu_{\alpha}$ are the conduction electron dispersions. Here $\alpha=L/R$ labels the two superconducting leads and $\bk$ and $\sigma$ are lead orbital and spin quantum numbers, respectively. The leads are assumed to have bandwidth $2D$ with $\xi_{k}\in[-D,D]$. The Land{\'{e}} $g$-factor in the superconductors is denoted by $g_{\mrce}$, the magnetic field by $B$, and the complex superconducting order parameters by $\Delta_{\alpha}=\Delta\e^{\i\phi_{\alpha}}$. Here we for simplicity take the two order parameters to be of equal magnitude. The normal-lead Hamiltonian is
\begin{equation}
H_{\mrN}=\sum_{\bk\s}(\ve_{N\bk\s}-\mu_{N})\cd_{N\bk\s}\can_{N\bk\s},
\end{equation}
with a bias voltage applied to this lead $\mu_{N}-\mu_{\alpha}=V$.  Assuming the level spacing in the quantum dot to be large, we model it by a single orbital:
\begin{equation}
H_{\D}=\sum_{\sigma=\up,\down}\ve_{d}\dd_{\sigma}\dan_{\sigma}+Un_{\up}n_{\down},
\quad \ve_{d\sigma}=\ve_d+\sigma g_\mri B/2,
\end{equation}
where $\dd_{\sigma}$ creates an electron in the orbital with spin $\sigma$, $U$ is the charging energy on the quantum dot, $\ve_{d}$ is the level position, which is controlled by the gate voltage, and $g_{\mri}$ is the $g$-factor in the dot (which can be different from $g_{\mrce}$).

The coupling between the leads and the dot is described by the tunneling Hamiltonian
\begin{equation}
H_{\T}=\sum_{\alpha k \sigma}\left(t_{\alpha}\cd_{\alpha k\sigma}\dan_{\sigma}+t_{\alpha}^{*}\dd_{\sigma}\can_{\alpha k\sigma}\right),
\end{equation}
where $t_{\alpha}$ denote the lead-dot tunneling amplitudes and the lead index $\alpha$ is extended to run through $L$, $R$, and $N$.

We focus on the odd-occupied spinful cotunneling regime well inside the corresponding Coulomb diamond, where
\begin{align}
2\pi\nu_{F}|t_{\alpha}|^{2}\equiv\Gamma_{\alpha}\ll -\varepsilon_{d},U+\varepsilon_{d},
\end{align}
with $\nu_F$ denoting the density of states at the Fermi level.

A standard Schrieffer-Wolff transformation~\cite{Salomaa1988,Schrieffer1966}  then leads to the following effective cotunneling (Kondo) model for the spin-$\tfrac{1}{2}$ coupled to the normal and two superconducting leads:
\begin{equation}\label{ham}
H=H_{\mrN}+H_{\LR}+H_{\mri,B}+H_{J}+H_{W},
\end{equation}
where the Zeeman term for the quantum dot spin reads
\begin{equation}
\label{hamB}H_{\mri,B}=g_{\mri}BS^{z},
\end{equation}
with $S^{i}$ denoting the spin operator on the dot.
The transformation is valid for $\Delta/U\ll 1$ and $|g_\mathrm{ce}-g_\mathrm{d}|B/U\ll1$.~\cite{Kirsanskas2014} For the polarized-spin approximation considered in the next section, the Zeeman term for the dot spin has no influence, but it will become important in \appendixname~\ref{sec:sbgstQS}
where the magnetic field dependence of the quantum corrections are discussed.

The \textit{exchange} cotunneling term reads
\begin{equation}
\label{hamJ}H_{J}=\sum_{\substack{i=x,y,z \\ \iap\s'\ia\s}} J_{\alpha'\alpha}^{{}}S^i\cd_{\iap\s'}\tau_{\sigma'\sigma}^i\can_{\ia\s},
\end{equation}
where $\tau^i$ denotes the Pauli matrices, and the \textit{potential} scattering term is
\begin{equation}
\label{hamW}
 H_{W}=\sum_{\substack{\iap,\ia,\s}}W_{\alpha'\alpha}^{{}}\cd_{\iap\s}\can_{\ia\s}.
\end{equation}
Here the exchange, and potential scattering amplitudes are given by
\begin{equation}\label{paramJW}
J_{\alpha\alpha'}=\frac{4}{1-x^2}\frac{t_{\alpha}t_{\alpha'}}{U}, \quad W_{\alpha\alpha'}=\frac{2x}{1-x^2}\frac{t_{\alpha}t_{\alpha'}}{U},
\end{equation}
where $x$ parameterizes the dimensionless gate voltage as
\begin{equation}\label{xdef}
x=1+\frac{2\ve_{d}}{U}.
\end{equation}
Note that the Anderson model always gives antiferromagnetic exchange $J>\abs{W}\geq 0$ inside the odd occupied diamond $x\in[-1,1]$ and that $W$ breaks particle-hole symmetry and therefore vanishes at the particle-hole symmetric point, $x=0$, defining the middle of the Coulomb diamond. We also define the following dimensionless coupling constants:
\begin{equation}\label{gdef}
\begin{aligned}
&g_{\alpha\alpha'}=\pi\nu_F J_{\alpha\alpha'}S,\quad &&g=g_{LL}+g_{RR},\\
&w_{\alpha\alpha'}=\pi\nu_F W_{\alpha\alpha'}, &&w=w_{LL}+w_{RR},
\end{aligned}
\end{equation}
to be used extensively below.

\section{Polarized-spin approximation}
\label{sec:polar}

We start by considering the case where the spin operator in Eq.~\eqref{hamJ} is treated as a classical variable with a fixed direction: $\mathbf{S}\approx S \hat{z}$. In this approximation, the problem is similar to the original problem considered by Yu, Shiba, and Rusinov,~\cite{Yu1965, Shiba1968, Rusinovfirst} but now with two superconductors having different phases. In \appendixname~\ref{sec:sbgstQS} we show that this approximation is justified when $g_\mathrm{d}B\gg g^2\Delta$. At zero field the excitation energies calculated within this approximation correspond to the correct result at weak coupling ($g\ll 1$) only after rescaling $g$ by a factor of 3. For now, we use the polarized-spin approximation in order to discuss the spectroscopy.

We start by diagonalizing the exchange, and potential scattering terms in $L/R$-space, omitting the coupling to the normal lead. This diagonalization is possible because they share the same matrix structure in $L/R$-lead space
\begin{align}
J_{\alpha\alpha'}=J\Theta_{\alpha\alpha'},\quad {\rm and}\quad
W_{\alpha\alpha'}=W\Theta_{\alpha\alpha'},
\end{align}
where
\begin{equation}
\Theta_{\alpha\alpha'}=\left(\begin{array}{cc}
\cos^2\theta & \sin\theta\cos\theta \\
\sin\theta\cos\theta & \sin^2\theta
\end{array}\right),
\end{equation}
with coupling asymmetry parameterized by an angle $\theta$ defined by
\begin{align}\label{thetadef}
(\cos\theta,\sin\theta)=(t_L,t_R)/t,\quad t=\sqrt{t_L^2+t_R^2}.
\end{align}
Notice that $\theta=\pi/4$ corresponds to symmetric coupling $t_{L}=t_{R}$. By means of a gauge transformation, the phases of the individual pairing potentials in the contacts, $\Delta _{\alpha }$, can be combined to a phase difference, $\phi=\phi_{L}-\phi_{R}$, appearing only in the scattering terms via the matrix $\Theta$:
\begin{equation}
\Theta _{\alpha \alpha^{\prime }}\rightarrow \left(
\begin{array}{cc}
\cos ^{2}\theta  & e^{i\phi/2}\sin \theta \cos \theta  \\
e^{-i\phi/2}\sin \theta \cos \theta  & \sin ^{2}\theta
\end{array}
\right).
\end{equation}
This matrix has eigenvalues $0$ and 1 with corresponding eigenvectors:
\begin{equation}
\mathbf{v}_{0}=\left(
\begin{array}{c}
e^{i\phi/2}\sin \theta  \\
-\cos \theta
\end{array}
\right) ,\quad \mathbf{v}_{1}=\left(
\begin{array}{c}
\cos \theta  \\
e^{-i\phi/2}\sin \theta
\end{array}
\right) ,
\end{equation}
and a unitary transformation that diagonalizes $J_{\alpha \alpha
^{\prime }}$ and $W_{\alpha \alpha ^{\prime }}$ is therefore achieved with
\begin{equation}
\mathbf{U}=\left(
\begin{array}{cc}
\cos \theta  & e^{i\phi/2}\sin \theta  \\
e^{-i\phi/2}\sin \theta  & -\cos \theta
\end{array}
\right),
\end{equation}
leading to the following transformed cotunneling terms:
\begin{equation}\label{Hscatrot}
H_{J}+H_{W}=\sum_{kk^{\prime }\sigma}(\sigma J S+W)
\tilde{c}_{1k\sigma }^{\dagger}\tilde{c}_{1k^{\prime }\sigma }^{{}},
\end{equation}
where the new operators $\tilde{c}_{0k}$ and $\tilde{c}_{1k}$ are defined as
\begin{equation} \label{dUc}
\left(
\begin{array}{c}
\tilde{c}_{1k}^{{}} \\
\tilde{c}_{0k}^{{}}
\end{array}
\right) =\mathbf{U}\left(
\begin{array}{c}
c_{Lk}^{{}} \\
c_{Rk}^{{}}
\end{array}
\right) .
\end{equation}
From expression \eqref{Hscatrot} it is evident that the channel corresponding to eigenvalue 0 does not couple to the cotunneling terms.

However, with finite phase the two channels are not independent and therefore the problem remains effectively a two-channel problem. In order to see this, we write lead Hamiltonians in Nambu space
\begin{equation}\label{eq:HNambu}
H_{0}=\frac{1}{2}\sum_{k\sigma}C_{k\sigma }^{\dagger }\left(
\begin{array}{cc}
\xi_{k}+\sigma g_\mrce B/2  & \Delta  \\
\Delta  & -\xi _{k}+\sigma g_\mrce B/2
\end{array}
\right) C_{k\sigma },
\end{equation}
with Nambu 4-spinors defined as
\begin{align}\label{eq:4Nambu}
C_{k\sigma }=(c_{Lk\sigma },c_{Rk\sigma },-\sigma c_{L-k-\sigma }^{\dagger },
-\sigma c_{R-k-\sigma }^{\dagger })^T.
 \end{align}
Notice that since $H_{0}$ is diagonal in lead space, a unit matrix in lead space is implied on each of the four matrix elements in~\eqref{eq:HNambu}.
After the rotation~\eqref{dUc}, which in Nambu space reads
\begin{equation}
\tilde{C}_{k\sigma }=\left(
\begin{array}{cc}
U & 0 \\
0 & U^{\ast }
\end{array}
\right) C_{k\sigma },\quad \tilde{C}_{k\sigma }^{\dagger }=C_{k\sigma }^{\dagger
}\left(
\begin{array}{cc}
U^{\dagger } & 0 \\
0 & U^{T}
\end{array}
\right),
\end{equation}
the lead Hamiltonian becomes
\begin{equation}\label{eq:HNambutrans}
H_{0}=\frac{1}{2}\sum_{k\sigma}\tilde{C}_{k\sigma }^{\dagger }\left(
\begin{array}{cc}
\xi _{k}+\sigma g_\mrce B/2 & \Delta P \\
\Delta P^{\dagger } & -\xi _{k}+\sigma g_\mrce B/2
\end{array}
\right) \tilde{C}_{k\sigma },
\end{equation}
where $P=U^{\dagger }U^{\ast }$, which evaluates to
\begin{equation}
P=\left(
\begin{array}{cc}
\cos ^{2}\theta +e^{i\phi }\sin ^{2}\theta  & -i\sin(2\theta)\sin(\phi/2)\\
-i\sin(2\theta)\sin(\phi/2)  & \cos^{2}\theta+e^{-i\phi}\sin^{2}\theta
\end{array}
\right) .
\end{equation}
From the expression for $P$, we now see that for $\phi \neq0$ the two channels are not independent, and the problem therefore remains a genuine two-channel scattering problem.

\subsection{Subgap spectrum from T-matrix}

We are interested in understanding the structure of the subgap states and therefore we study the $T$-matrix for spin-$\sigma$ given by
\begin{equation}\label{Tmatrix}
T_{\sigma }^{R}(\omega )=V_{\sigma}\left[ 1-G_{0}^{R}(\omega-\sigma g_\mrce B/2 )V_{\sigma}\right]
^{-1},
\end{equation}
with the diagonalized exchange, and potential scattering terms extended to Nambu space as
\begin{equation}\label{Vdef}
V_{\sigma}=\left(
\begin{array}{cccc}
\sigma J S+W & 0 & 0 & 0 \\
0 & 0 & 0 & 0 \\
0 & 0 & \sigma J S-W & 0 \\
0 & 0 & 0 & 0
\end{array}
\right),
\end{equation}
whereby
\begin{align}
H_{J}+H_{W}=\frac{1}{2}\sum_{kk'\sigma}\tilde{C}^{\dagger}_{k\sigma}V_{\sigma}
\tilde{C}_{k'\sigma}.
\end{align}
The local Green's function corresponding to~\eqref{eq:HNambutrans} is found as
\begin{align}\label{G0def}
G_{0}^{R}(\omega)=&\sum_{k}\left[\omega-\left(
\begin{array}{cc}
\xi _{k} & \Delta P \\
\Delta P^{\dagger } & -\xi _{k}
\end{array}%
\right) \right] ^{-1}\\
=&-\frac{\pi \nu_{F}}{\sqrt{\Delta ^{2}-\omega
^{2}}}\left(
\begin{array}{cccc}
\omega  & 0 & a\Delta & -ic\Delta \\
0 & \omega  & -ic\Delta & a^{\ast }\Delta \\
a^{\ast}\Delta & ic\Delta & \omega  & 0 \\
ic\Delta & a\Delta & 0 & \omega
\end{array}
\right),\nonumber
\end{align}
where the $k$-integration was performed assuming a constant density of states $\nu_F$ and assuming that $\abs{\omega}<\Delta\ll D$. The lead asymmetry and the phase-difference are encoded in the dimensionless coefficients
\begin{align}
a&=\cos ^{2}\theta +e^{i\phi }\sin ^{2}\theta,\\
c&=\sin(2\theta)\sin(\phi/2).
\end{align}
The condition for poles in $T^{R}_{\sigma}$ is
\begin{equation}\label{Gpoles}
\det\left[1-G_{0\sigma}^{R}(E-\sigma g_\mrce B/2)V_{\sigma}\right]=0,
\end{equation}
and after some algebra one finds two roots for each spin, corresponding to YSR states at energies
\begin{equation}\label{esosaE}
\begin{aligned}
E_{\pm,\s}&=\sigma g_\mrce B/2-\frac{\s c_{\pm}\absDelta}{\sqrt{(1+u)^2+4g^2}}
\Big[(1+u)(1+\chi u)\\
&+2g^2\pm2g\sqrt{g^2+u(1-\chi)(1+\chi u)}\Big]^{1/2},
\end{aligned}
\end{equation}
with the following shorthand notation
\begin{equation}\label{chi}
\begin{aligned}
\chi&=1-\sin^2(2\theta)\sin^2(\phi/2),\\
u&=w^2-g^2,\\
c_{-}&=\sgn(1+\chi u),\\
c_{+}&=1.
\end{aligned}
\end{equation}
Here we have assumed $g>0$; the corresponding solutions for $g<0$ simply have opposite spins, and hence are given by $E_{\pm,-\s}$.
\begin{figure}[t!]
\includegraphics[width=0.5\textwidth]{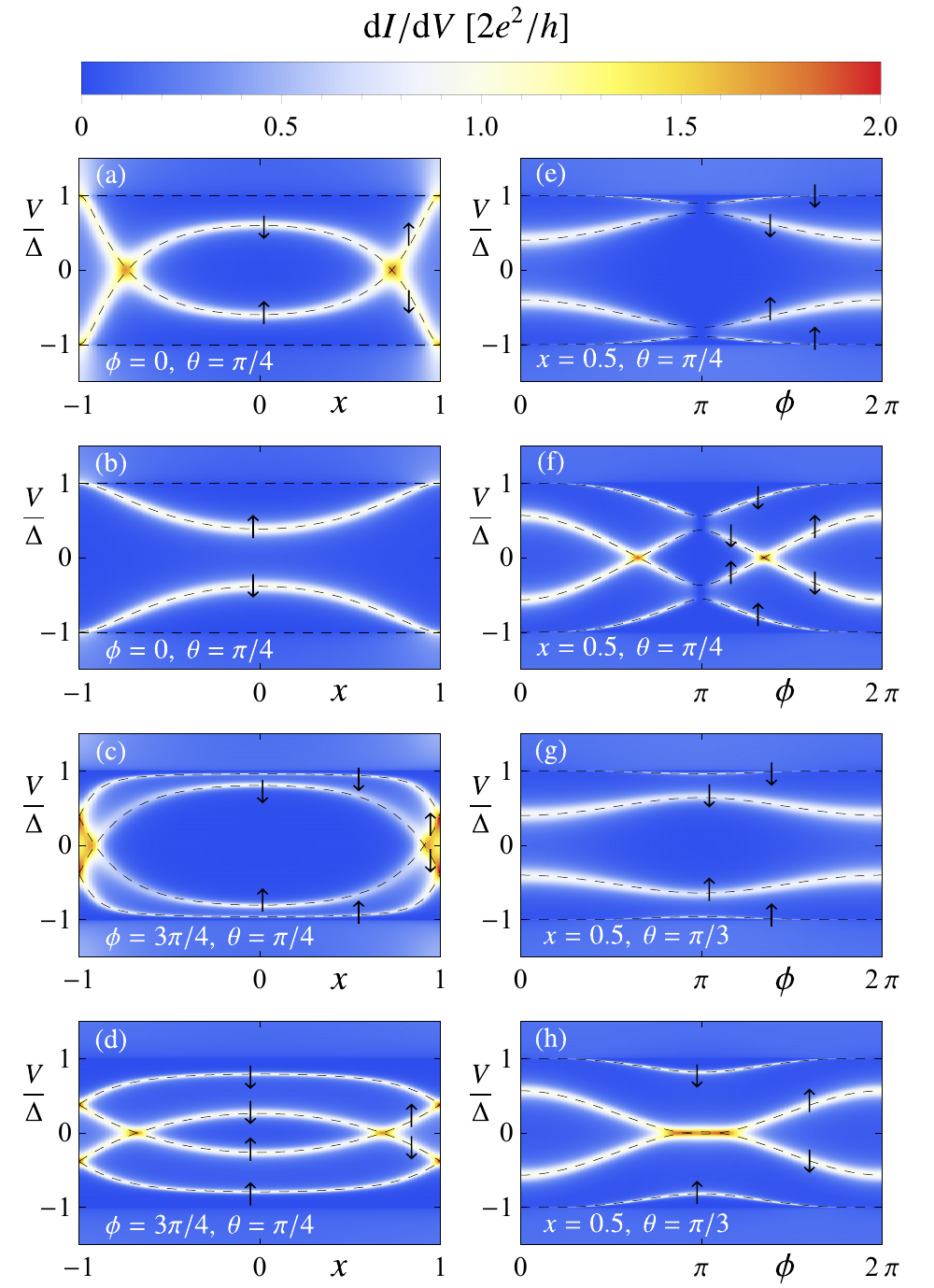}
\caption{\label{fig3}(Color online)
The figure shows the YSR state energies $V=E_{\pm,\sigma}$ given in Eq.~\eqref{esosaE} as dashed lines. The YSR energies are plotted as a function of dimensionless gate voltage $x=1+2\varepsilon_d/U$ in (a)-(d), and as function of phase difference in (e)-(h). $x=0$ corresponds to the middle of the odd diamond and $x=\pm 1$ are the charge degeneracy points, see Eqs.~\eqref{paramJW} and \eqref{xdef}.  The lines are overlayed on contour plots of the differential conductance (see \figurename~\ref{fig1} and Eq.~\eqref{gnssmtr}) using $g_{NN}=0.1$ and the parametrization of the couplings given in Eq.~\eqref{normcoupparam}. The arrows indicate the spin of the excited states. The exchange coupling at the middle of the diamond is set to $g(x\!=\!0)=0.5$ in figures (a), (c), (e), and (g) and to $g(x\!=\!0)=1.5$ in (b), (d), (f), and (h). In (a), the YSR states are seen to cross zero energy when the effective exchange coupling increases as $x$ is changed from 0 to 1 towards the charge degeneracy points. (b) With a stronger coupling, the eye-shaped feature in the middle becomes smaller, and for strong enough coupling it disappears. (c) With finite phase difference, two new bound states appear. (d) A finite phase difference reduces the effective exchange coupling and restores the eye-shaped crossings. The detailed phase evolution is shown in (e) and (f) for a cut in gate voltage corresponding to $x=0.5$. While panels (a)-(f) are for a symmetrically coupled junction ($\theta=\pi/4$), (g) and (h) show the phase dependence for an asymmetrically coupled junction ($\theta=\pi/3$). Note that the finite-bias degeneracy at $\phi=\pi$ is lifted by the asymmetry.}
\end{figure}

In the rest of this section we examine the dependence of the subgap states on the coupling strengths and the phase difference. For zero phase difference, $\phi=0$, and antiferromagnetic coupling, $g>0$, Eq.~\eqref{esosaE} gives the following subgap energies:
\begin{equation}\label{Eysr}
E_{\sigma}=-\sigma\absDelta\frac{1+w^2-g^2}{\sqrt{(1+w^2-g^2)^2+4g^2}}, \quad g>0.
\end{equation}
This is the result obtained by Yu, Shiba and Rusinov.~\cite{Yu1965,Shiba1968,Rusinovfirst}  In the case when there is no exchange coupling, i.e., $g=0$, Eq.~\eqref{esosaE} yields the usual expression for the Andreev bound state:~\cite{Haberkorn1978,Zaitsev,Arnold1985,Furusaki1990,Beenakker1991}
\begin{equation}
E^{0}_{\pm,\sigma}=\pm\sigma\absDelta\sqrt{1-\tau\sin^2\frac{\phi}{2}},
\quad \tau=\frac{w^2\sin^2(2\theta)}{1+w^2},
\end{equation}
where $\tau$ denotes the normal state transmission of the junction.

The energies of the four bound states are plotted as dashed curves against gate voltage and phase difference for different cases in \figurename~\ref{fig3}. The color scale refers to the differential conductance for the system with an added normal lead (see next section). In panels (a) and (b) we plot the bound-state energies for zero phase difference and weak, and strong coupling to the superconductors, respectively. The difference between (a) and (b) can be understood from the position of the YSR states at $x=0$ (where $W=0$) and $\phi=0$:
\begin{equation}\label{YSRx0}
  E_\sigma= -\sigma\frac{1-g^2}{1+g^2}.
\end{equation}
For (a), where $g=0.5$ at $x=0$, the upper YSR state moves down and crosses zero as $x$ approaches the charge degeneracy points at which $g$ diverges. In contrast, in (b) where $g=1.5$ at $x=0$, the YSR state has already crossed zero at $x=0$ due to stronger coupling. Interestingly, the zero energy crossings
correspond to a change of parity of the ground state, cf. the discussion in Ref.~\onlinecite{Balatsky2006}. Adding now a finite phase difference, panels (c) and (d) reveal the 2-channel nature of the problem, with two bound states above and below the zero energy.

In panels (e)-(h), we plot the dispersion of the bound-state with phase difference. In general, a finite phase difference is seen to to shift the value of the coupling at which the energy levels cross zero to higher values. In the plots, $\theta=\pi/4$ corresponds to symmetric coupling, $t_L=t_R$, see Eq.~\eqref{thetadef}. For the particle-hole symmetric point $x=0$ and symmetric coupling $\theta=\pi/4$, we see that the two excitations are degenerate at $\phi=\pi$.

It is interesting to compare the above features with the experimental results by Chang \textit{et al.}~\cite{Chang2012} They show conductance plots similar to \figurename~\ref{fig3} (a)-(b) for three different ranges of back-gate voltage, corresponding to different devices with either strong, or weak coupling, as well as one device right at the transition where the YSR states touch at zero energy in the middle of the diamond at $x=0$. Furthermore, Chang \textit{et al.} show the phase dependence for different fixed gate voltages similar to \figurename~\ref{fig3} (e) and (f) and with similar qualitative features: only weak gate dependence in the weakly coupled device, and a restoring of the zero-energy crossings at finite phase difference for the stronger coupled devices.

We have seen that a finite phase difference results in two subgap states at positive energy. Interestingly, the same situation occurs for a magnetic impurity coupled to an $s_\pm$ superconductor, i.e., a superconductor with two bands, where the pairing potentials have different signs in the two bands.~\cite{matsumoto2009} Note that if the two superconductors have different pairing potentials, but with same sign, there is only one bound state within the smallest of the two gaps.

\subsection{\label{sec:condnmp}
Conductance to the normal-metal tunnel probe}

\begin{figure}[ht!]
\begin{center}
\includegraphics[width=0.475\textwidth]{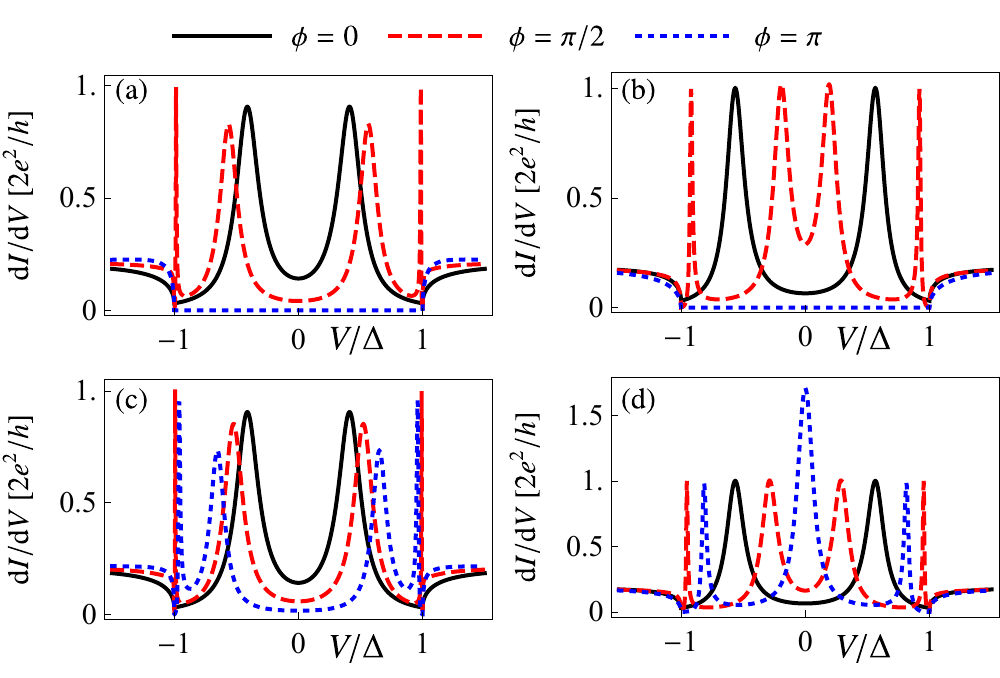}
\end{center}
\caption{\label{fig4}
(Color online) Differential conductance $\mathrm{d}I/\mathrm{d}V$ in units of $2e^2/h$ between the normal electrode and the superconducting leads, see \figurename~\ref{fig1}. The different lines correspond to constant phase cuts ($\phi=0,\pi/2,\pi$) in \figurename~\ref{fig3}(e)-(h), as indicated at the top of the figure. The left panels (a) and (c) are for weak coupling, ($g(x\!=\!0)=0.5$), while the right panels (b) and (d) are for stronger coupling, $g(x\!=\!0)=1.5$. The top panels (a) and (b) are for symmetric coupling $\theta=\pi/4$, while the bottom panels (c) and (d) are for asymmetric coupling $\theta=\pi/3$. In all cases the gate voltage is set to the particle-hole symmetric point, $x=0$, and the coupling to the normal lead was chosen to be $g_{NN}=0.1$.}
\end{figure}
The YSR subgap states derived above can be observed by means of tunneling spectroscopy from a normal metal lead, as illustrated in \figurename~\ref{fig1} and experimentally carried out in Refs.~\onlinecite{Pillet2010, Chang2012}. We assume the $N$-dot tunneling rate to be larger than any relaxation rate between the YSR state and the quasiparticle continuum in the superconductors, thus ruling out the single-electron tunneling currents which were recently demonstrated to be important for analyzing scanning tunneling spectroscopy of Mn adatoms on a Pb(111) surface.~\cite{Ruby2015} In this case, the current is carried exclusively by elastic Andreev reflections and the differential conductance between the normal lead and the superconducting region can be obtained from the $S$-matrix in the normal lead as~\cite{Blonder1982,Lambert1991,Takane1992,Beenakkerreview} (at zero temperature)
\begin{equation}\label{gnssmtr}
\frac{\mathrm{d}I}{\mathrm{d}V}=\frac{e^2}{h}\Tr[2-S_{ee}^{\phantom{\dag}}S_{ee}^{\dag}+S_{he}^{\phantom{\dag}}S_{he}^{\dag}],
\end{equation}
where $S_{ee}$ and $S_{he}$ are the scattering amplitudes at energy $eV$ for an incident electron to be reflected as an electron or a hole, respectively. The amplitudes $S_{ee}$ and $S_{he}$ can be obtained from the retarded $T$-matrix in Eq.~\eqref{Tmatrix}, which we write here as (setting $B=0$ for simplicity)
\begin{subequations}
\begin{align}
&\TTT^R(\omega)=\VVV[1-\GGG_{0}^R(\omega)\VVV]^{-1},\\
&\VVV=S \JJJ \sigma^{3}+\WWW\tau^{3},
\end{align}
\end{subequations}
where $\tau^{i}$ and $\sigma^{i}$ are Pauli matrices representing particle-hole and spin space, respectively, and $\JJJ$ and $\WWW$ are 3$\times3$ coupling matrices in lead space, with elements $J_{\alpha\alpha'}$ and  $W_{\alpha\alpha'}$ for $\alpha,\alpha'=L,R,N$.

The unperturbed momentum-summed Green's function is a diagonal matrix in lead space,
where for the superconducting leads $\alpha=L,R$ it is
\begin{align}
\label{gnotksum}
G^{R}_{0,\alpha\alpha}(\omega)=&-\pi\nu_F\frac{ (\omega+\i\eta)-\Delta_{\alpha}(\tau^{1}\cos\phi-\tau^{2}\sin\phi)}
{\sqrt{\abs{\Delta_{\alpha}}^2-(\omega+\i\eta)^2}},\notag
\end{align}
where $\eta$ is a positive infinitesimal. For the normal lead, the Green's function simplifies to $G^{R}_{0,NN}(\omega)=-i\nu_F\pi$, assuming the normal-lead density of states $\nu_{F}$ to be the same as for the two superconductors, since any difference can be absorbed into the tunneling matrix elements. The $S$-matrix is now expressed through the $T$-matrix as
\begin{equation}
\SSS(\omega)=1-2\pi\i\nu_{F}\TTT(\omega),
\end{equation}
and the amplitudes $S_{ee}$ and $S_{he}$ are found as submatrices of $\SSS$ with $\alpha'=\alpha=N$ and component (2,1) in electron-hole space for $S_{he}$ and (1,1) for $S_{ee}$. The calculation can be carried out analytically, but is quite lengthy. In \appendixname~\ref{sec:appendixAB}, we provide an analytical expression for $\mathrm{d}I/\mathrm{d}V$ when the coupling to the superconducting leads is symmetric, $\theta=\pi/4$, see Eq.~\eqref{gnsanl}.

In \figurename~\ref{fig4} we present the differential conductance when the subgap states are probed by the normal lead. The traces correspond to vertical cuts in \figurename~\ref{fig3}. The coupling is chosen to be $g_{NN}=0.1$: weak enough to resolve the YSR states as distinguishable conductance peaks, and large enough to actually see them. For the cases with two bound states, both states give rise to peaks in $\mathrm{d}I/\mathrm{d}V$, but with different widths. Assuming $\phi=0$, $\theta=\pi/4$, $x=0$, and $g_{NN}\ll g_{LL}=g_{RR}$, the width of the subgap conductance peak can be found from Eq.~\eqref{gnsanl} to be proportional to $g_{NN}\Delta (1-E_{\sigma}^{2}/\Delta^{2})$, with $E_{\sigma}$ from~\eqref{Eysr}, implying very sharp peaks close to the gap-edges, and an overall scale set by the width $g_{NN}\Delta$ of a deep YSR state. When the voltage is resonant with subgap states the differential conductance is close to $2e^2/h$, except when the two subgap states are degenerate at zero energy, in which case they add up to exactly $4e^2/h$ (cf. \figurename~\ref{fig3}~(h)), or for a symmetric junction ($\theta=\pi/4$) when they are degenerate at finite energy for $\phi=\pi$, where the conductance is exactly zero (cf. \figurename~\ref{fig3}~(e-f)).

\subsection{\label{sec:scrt}Supercurrent}

\begin{figure}
\begin{center}
\includegraphics[width=0.9\columnwidth]{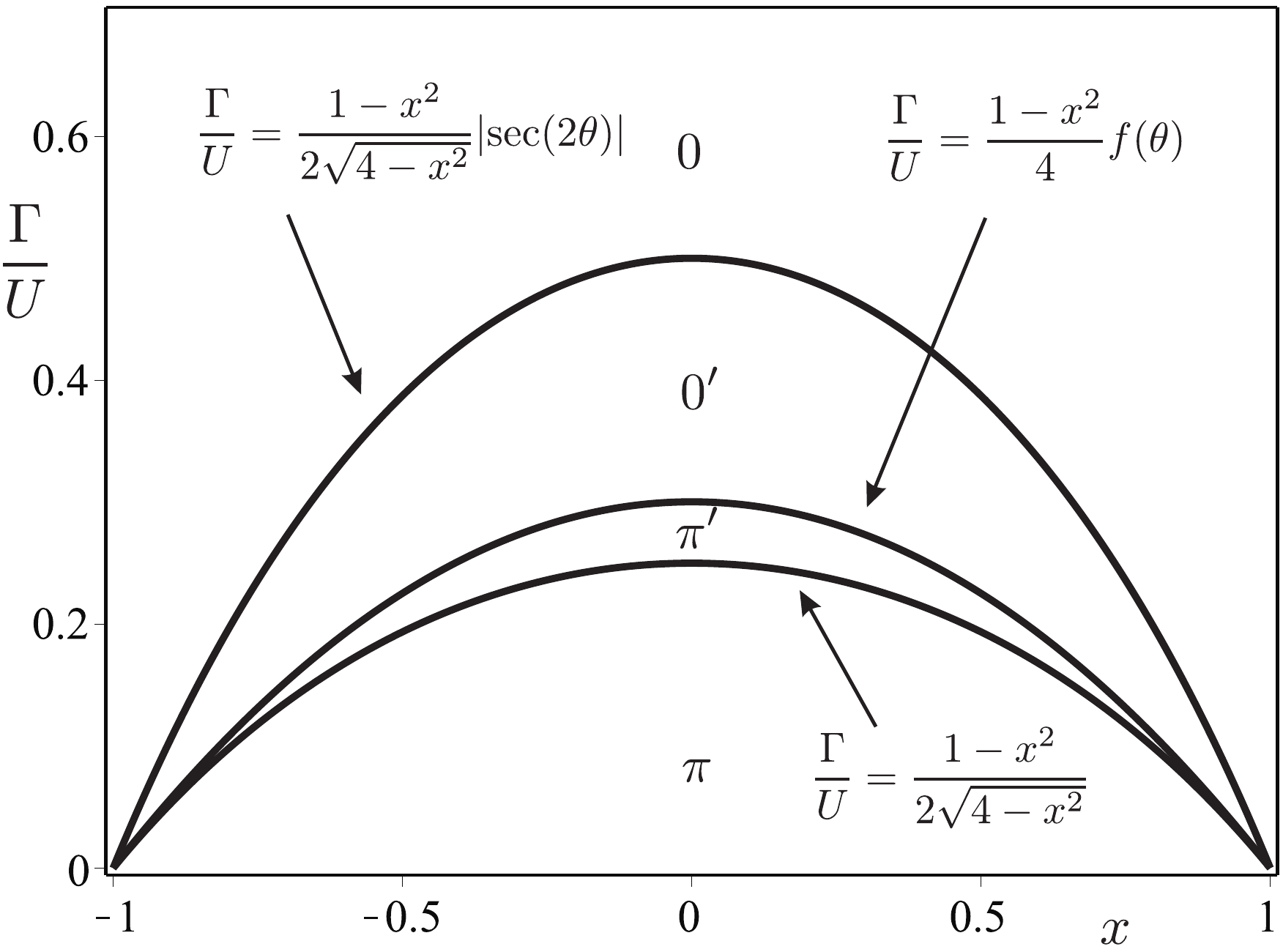}
\caption{\label{fig5}
Phase diagram with couplings parameterized by Eq.~\eqref{paramJW}, in terms of  $\Gamma=\nu_F(t_{L}^2+t_{R}^2)$, and the dimensionless level position (gate voltage) $x=1+2\ve_{d}/U$. The phase boundaries are independent of $\Delta$, and their dependences on $x$ and $\theta$ are indicated, with the function $f(\theta)$ defined as
$f(\theta)=[1/2\big\{\sin^2(2\theta)+\sqrt{4+\sin^4(2\theta)}\big\}]^{1/2}$. In this diagram the coupling asymmetry parameter was chosen to $\theta=\pi/3$, whereas for a symmetric junction with $\theta=\pi/4$, the 0'-0-boundary will never be reached, consistent with earlier results on the Anderson model.~\cite{Rozhkov1999,Tanaka2007,Zonda2015}}
\end{center}
\end{figure}

\begin{figure}
\begin{center}
\includegraphics[width=0.9\columnwidth]{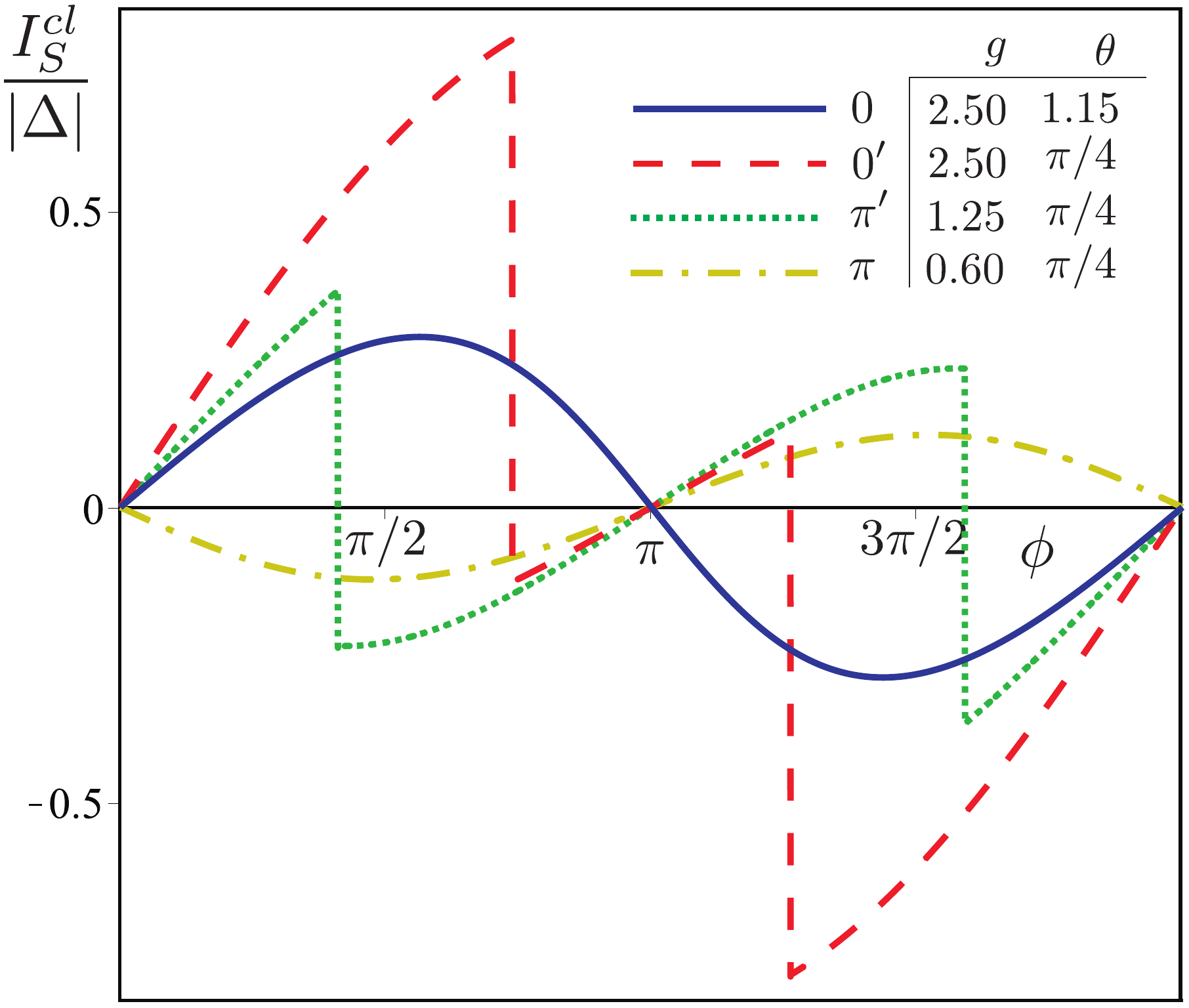}
\caption{\label{fig6}(Color online) Supercurrent vs. phase difference at the particle-hole symmetric point, $x=0$, for exemplary values of $g$ and $\theta$ (see inset) corresponding to the four different kinds of Josephson junctions. Similar results were first obtained in Refs.~\onlinecite{Rozhkov1999,Choi2004} and experimentally demonstrated in Ref.~\onlinecite{Delagrange2015}.}
\end{center}
\end{figure}

At zero temperature, the supercurrent can be found as the derivative of the ground-state energy, $E_{\mathrm{GS}}$, with respect to the phase difference between the two superconducting leads:~\cite{Caianiello1964}
\begin{equation}\label{degsdphi}
I_{S}(\phi)=2\frac{\pd E_{\mathrm{GS}}}{\pd \phi}.
\end{equation}
If the ground-state energy has a single minimum either at $\phi=0$ or at $\phi=\pi$, the junction is classified respectively as a $0$-junction or a $\pi$-junction. In the former ($0$) case, the supercurrent is a continuous function of $\phi$ with positive slope at $\phi=0$ and negative slope at $\phi=\pi$. In the latter ($\pi$) case, the supercurrent has negative slope at $0$ and a positive slope at $\pi$. In cases with minima both at $\phi=0$ and at $\phi=\pi$, a $0'$- or a $\pi'$-junction refers to the global minimum being respectively at $0$ or $\pi$, and the supercurrent is now a discontinuous function (with one discontinuity in the interval $\phi\in[0,\pi]$ and one in the interval $\phi\in[\pi,2\pi]$) with positive slopes both at $\phi=0$ and at $\phi=\pi$.

Within the polarized-spin approximation, we find the continuum to be independent of the phase difference in the limit of infinite bandwidth. This is is akin to a so-called {\it short} junction~\cite{Beenakker1991} and implies that the supercurrent can be obtained from the subgap excitation spectrum alone, which according to the standard Bogoliubov-de Gennes formalism is given by
\begin{equation}\label{EGSeq}
E_{\mathrm{GS}}=-\frac{1}{2}\left(\abs{E_{+}}+\abs{E_{-}}\right)+\mathrm{const}.
\end{equation}
The resulting phase diagram as a function of the dimensionless gate voltage $x$, and the lead-dot tunneling rate, $\Gamma=\pi\nu_F(t_{L}^2+t_{R}^2)$, is depicted in \figurename~\ref{fig5}, and examples of supercurrent in the different phases are shown in \figurename~\ref{fig6}. Only for small coupling constants do we have a sinusoidal current-phase relations:
\begin{equation}\label{ISclas}
I_{S}\approx(\absDelta/2)\sin^2(2\theta)\left(w^2-g^2\right)\sin\phi.
\end{equation}

In some cases the supercurrent is discontinuous, which is directly related to the subgap states crossing zero energy and changing their spin (as is evident from Eq.~\eqref{EGSeq}). Similar behavior of the supercurrent has been demonstrated experimentally in Ref.~\onlinecite{Delagrange2015}, where a thermally smoothened discontinuity was observed at a phase difference $\phi_{c}$ whose dependence on gate gate voltage was shown to comply well with the formula
\begin{align}
\phi_{c}=2\arccos[\sqrt{\gamma-(x/h)^2}].
\end{align}
This functional dependence on the dimensionless gate voltage, $x$, was derived in the atomic limit ($\Delta\rightarrow\infty$) in Ref.~\onlinecite{Karrasch2008}. However, in the experiment $\Delta\sim T_{K}$, and $\gamma$ and $h$ were therefore used as free fitting parameters. Moreover, the data was also shown to match the result of a quantum Monte Carlo calculation for the corresponding Anderson model. We note here that the same functional behavior follows directly from Eqs.~\eqref{esosaE},~\eqref{degsdphi}, and~\eqref{EGSeq}, with coefficients $\gamma$ and $h$ defined as
\begin{align}
\gamma=\frac{1}{h^2}-\frac{1}{\tan^2(2\theta)},\quad
h=g\sin(2\theta),
\end{align}
which allows extracting the coupling asymmetry and the dimensionless exchange coupling.

\section{Quantum mechanical treatment of the spin degree of freedom}
\label{sec:Q}

Here we discuss the results for the fully quantum mechanical description, where unlike for the polarized-spin approximation an exact solution is not possible. Therefore we resort to perturbation in the cotunneling couplings $g$ and $w$, which is valid when the Kondo temperature is much smaller than the superconducting gap $T_{K}\ll\Delta$ and the ground state is a doublet. Details of the calculations can be found in \appendixname~\ref{sec:sbgstQS}, and here we simply summarize the main findings and contrast them to the results for the polarized spin found above.

In \appendixname~\ref{sec:qspm}, we diagonalize an effective low-energy single-quasiparticle Hamiltonian. As in the spin-polarized case, we find two subgap excitation energies when the two superconductors have a phase difference. The transition energy from the ground-state doublet to the excited subgap singlet perturbatively matches the second order expansion (in $g_{\alpha\alpha'}$) of the excitation energy obtained from the polarized-spin approximation for $B=0$ and $S=\frac{1}{2}$, when replacing $g_{\alpha\alpha'}\rightarrow 3g_{\alpha\alpha'}$ in \eqref{esosaE}.

In \appendixname~\ref{sec:qbfield}, we show how the subgap excitation energies at weak coupling, $g\ll 1$, are shifted in energy by an external magnetic field. Assuming that the field is screened in the superconducting leads, $g_{\mrce}\approx 0$, the transition energy from the ground-state doublet to the excited-state singlet is found to approach the polarized-spin result in the limit of high magnetic field, $B\gg g^{2}\abs{\Delta}$ (see \figurename~\ref{fig7}).

\appendixname~\ref{sec:qbeyond} goes beyond the low-energy single-quasiparticle Hamiltonian and provides a calculation of the leading order ($g_{\alpha\alpha'}^{2}$) correction to the ground-state energy. With reference to the three-quasiparticle Yosida wavefunction ansatz analyzed in Ref.~\onlinecite{Kirsanskas2014}, we argue why this shift of ground-state energy does not modify the excitation energy found in \appendixname~\ref{sec:qspm}.

Finally, this second order shift in ground-state energy is used in \appendixname~\ref{sec:qscrt} to calculate the supercurrent in the presence of a finite magentic field via formula~\eqref{degsdphi}. In the perturbative regime, the functional dependence of the supercurrent on both $\phi$ and $\theta$ is similar with, and without the polarized-spin approximation, but again, only in the large-field limit do they match exactly.

Summarizing theses findings, we find qualitative agreement between the results found within the polarized-spin approximation and the perturbative results for the full quantum mechanical treatment. A non-perturbative calculation of the YSR spectrum and the supercurrent beyond the polarized-spin approximation could be found by means of a numerical renormalization group calculation of the $T$-matrix, whereas the non-linear Andreev conductance would require an average of the current operator.

\section{\label{sec:concl}Conclusions}

In summary, we have determined the phase-dispersion of the Yu-Shiba-Rusinov states induced by a spinful Coulomb blockaded quantum dot coupled to two phase-biased superconducting leads. At finite phase difference, two channels are involved in the screening of the dot spin. Consequently, the phase-biased system exhibits two, instead of one, YSR states, one of which merges with the continuum at the gap edge for zero phase difference. We have shown how the corresponding subgap excitation spectrum is modified by coupling asymmetry and potential scattering, and established that the phase difference generally shifts the parity transition, and the accompanying sign change in the supercurrent, to larger values of the exchange coupling.

We have solved the problem exactly in the spin-polarized approximation, and perturbatively in the fluctuating quantum case  (cf.~\appendixname~\ref{sec:sbgstQS}), and obtained a closed analytical expression for the subgap excitation energy [Eq.~\eqref{esosaE}], which extends Rusinov's result~\cite{Rusinovfirst} to a situation with two superconductors at a finite phase difference and arbitrary tunnel couplings.

The YSR bound states can for example be observed by a normal metal tunnel probe connected to the dot, which would also give information about their spectral weight. We have therefore calculated the differential conductance in such a setup. These results should provide a valuable basis for a more detailed analysis of future experiments like Refs.~\onlinecite{Chang2012, Pillet2010}.

\begin{acknowledgments}
We thank W. Chang, R. Delagrange, R. Deblock and I. Weymann for useful discussions. The research was supported by the Danish Council for Independent Research, Natural Sciences and by the Center for Quantum Devices funded by the Danish National Research Foundations. We also acknowledge support from the Simons Foundation and the BIKURA (FIRST) program of the Israel Science Foundation (M.G.), and US NSF DMR-1206612 (L.G.).
\end{acknowledgments}

\appendix

\section{\label{sec:appendixAB} Explicit formula for the symmetric-coupling conductance}

In the case of symmetric coupling of the quantum dot to the two superconductors ($\theta=\pi/4$), we can derive a simple closed-form expression for the differential conductance $\mathrm{d}I/\mathrm{d}V$. For subgap conductance ($\abs{V}<\Delta$) we obtain
\begin{widetext}
\begin{subequations}\label{gnsanl}
\begin{equation}
\frac{\mathrm{d}I}{\mathrm{d}V}=\frac{4t_{+}t_{-}f^2\cos^2\frac{\phi}{2}\times2e^2/h}
{[1 + (t_{+}t_{-}-r_{+}r_{-})(1-f^2\sin^{2}\frac{\phi}{2}) + (r_{+}+r_{-})fv]^2
+[(r_{+}t_{-}+r_{-}t_{+})(1-f^2\sin^{2}\frac{\phi}{2}) - (t_{+}+t_{-})fv]^2}+(v\rightarrow-v),
\end{equation}
and for continuum conductance ($\abs{V}>\Delta$) we get
\begin{equation}
\frac{\mathrm{d}I}{\mathrm{d}V}=\frac{2f\abs{v}[ (r_{+}^2t_{-}+r_{-}^2t_{+})(1+f^2\sin^2\frac{\phi}{2}) + (t_{+}+t_{-})\{1+t_{+}t_{-}(1+f^2\sin^2\frac{\phi}{2})\}+4t_{+}t_{-}f\abs{v} ]\times2e^2/h}
{[1 + (t_{+}t_{-}-r_{+}r_{-})(1+f^2\sin^2\frac{\phi}{2}) + (t_{+}+t_{-})f\abs{v}]^2
+[(r_{+}t_{-}+r_{-}t_{+})(1+f^2\sin^2\frac{\phi}{2}) + (r_{+}+r_{-})f\abs{v}]^2}.
\end{equation}
\end{subequations}
\end{widetext}
with $v=V/\Delta$ and $f(V)=\Delta/\sqrt{|\Delta^2-V^2|}$, and where transmission, and reflection amplitudes for incident electrons in the superconductors have been identified as
\begin{align}
t_{\pm}&=\frac{2(g_{NS}\pm w_{NS})^2}{1+(g_{NN}\pm w_{NN})^2},\\
r_{\pm}&=g\pm w-(g_{NN}\pm w_{NN})t_{\pm}.
\end{align}
where $g_{NS}\equiv g_{NL}=g_{NR}$ and $w_{NS}\equiv w_{NL}=w_{NR}$. Similar expressions to Eq.~\eqref{gnsanl} were derived for polarized spin coupled to a single superconductor\cite{Koerting2010} and unconventional superconductor junctions containing subgap states.\cite{Tanaka1995} For calculations of $\mathrm{d}I/\mathrm{d}V$ in Fig.~3 when there is finite coupling asymmetry, the couplings involving the normal lead are parametrized as
\begin{equation}\label{normcoupparam}
\begin{aligned}
&g_{NL}=\sqrt{g_{NN}\,g}\cos\theta, \quad &&g_{NR}=\sqrt{g_{NN}\,g}\sin\theta,\\
&w_{N\alpha}=g_{N\alpha}w/g, \quad &&w_{NN}=g_{NN}w/g.
\end{aligned}
\end{equation}

\section{\label{sec:sbgstQS}Subgap states beyond the polarized-spin approximation}

In this appendix, we determine the subgap states induced by the dot spin without doing the polarized-spin approximation applied in \sectionname~\ref{sec:polar}. The presence of the spin-flip terms prohibits a solution along the lines in the main part of the paper and we therefore revert to a perturbative treatment, valid for small dimensionless couplings. This problem was considered already by Soda, Matsuura and Nagaoka,~\cite{Soda1967} using Yosida's wave-function ansatz.~\cite{Yosida1966} Since the calculation using the ansatz is rather technical, it is more instructive to calculate the subgap spectrum, using an effective single-quasiparticle model for the case with no potential scattering at $x=0$.

First, the BCS leads are diagonalized by the Bogoliubov transformation
\begin{equation}\label{ceoioBqp}
\can_{\ia\s}=u_{\ia}\gaman_{\ia\s}+\s v_{\ia}\e^{\i\phi_{\alpha}}\gamd_{\alpha,-\bk\bar{\s}},
\end{equation}
where
\begin{equation}
u_{\alpha\bk}=\sqrt{\frac{1}{2}\left(1+\frac{\xi_{\bk}}{E_{\ia}}\right)},
\quad v_{\ia}=\sqrt{\frac{1}{2}\left(1-\frac{\xi_{\bk}}{E_{\ia}}\right)}.
\end{equation}
In terms of the Bogoliubov quasiparticle operators $\gamma_{\ia\s}$, the lead Hamiltonian (\ref{hamLR}) reads
\begin{equation}
\label{hamLR2}H_{\LR}=\sum_{\ia\s}E_{\ia\s}\gamd_{\ia\s}\gaman_{\ia\s},
\end{equation}
with eigenenergies
\begin{equation}
E_{\ia\s}=E_{\ia}+\s \frac{g_{\mrce}B}{2},\quad E_{\ia}=\sqrt{\xi_{\bk}^2+\abs{\Delta_{\alpha}}^2}.
\end{equation}

\subsection{\label{sec:qspm}Effective single-quasiparticle model}

Next, we express the exchange Hamiltonian $H_J$ in terms of the Bogoliubov operators. In accordance with the leading order term in Yosida's ansatz, we neglect all pairing-like terms, $\gamd_{a'\up}\gamd_{a\down}$ and $\gaman_{a'\up}\gaman_{a\down}$ in the exchange Hamiltonian and obtain the following low-energy effective model:
\begin{equation}\label{shamJ}
\begin{aligned}
   H_J&\approx\frac{1}{2}\sum_{\substack{\iap\ia}}\left(1+\e^{\i(\phi_{\alpha'}-\phi_{\alpha})}\right)J_{\alpha'\alpha}\\
      &\phantom{......}\times\Big[S^{z}\big(\gamd_{\iap\up}\gaman_{\ia\up}-\gamd_{\iap\down}\gaman_{\ia\down}\big)\\
      &\phantom{.........}+S^{+}\gamd_{\iap\down}\gaman_{\ia\up}+S^{-}\gamd_{\iap\up}\gaman_{\ia\down}\Big]\\
         &=\sum_{\bk'\bk}\psi^{\dag}_{\bk'}\m{M}\psi_{\bk},
 \end{aligned}
\end{equation}
where we have set $u_{\bk}\approx v_{\bk}\approx\tfrac{1}{2}$ since their energy dependence only matters for higher order corrections to the subgap excitation energies. The last line in \eqref{shamJ} is expressed in terms of the conduction electron 4-spinor
\begin{equation}
\psi^{\dag}_{\bk}=(\gamd_{L\bk\up},\gamd_{R\bk\up},\gamd_{L\bk\down},\gamd_{R\bk\down}),
\end{equation}
and the matrix
\begin{equation}\label{spinleadtp}
\begin{aligned}
\m{M}&=\begin{pmatrix}
S^{z} & S^{+} \\
S^{-} & -S^{z}
\end{pmatrix}\otimes
\begin{pmatrix}
J_{LL} & J_{LR}\frac{1+\e^{\i\phi}}{2} \\
J_{LR}^{*}\frac{1+\e^{-\i\phi}}{2} & J_{RR}
\end{pmatrix}\\
&=\m{M}_{s}\otimes\m{M}_{l},
\end{aligned}
\end{equation}
in which $\m{M}_{s}$ operates in spin, and $\m{M}_{l}$ in lead space.

The Hamiltonian \eqref{shamJ} is written in the excitation basis with $\gaman_{\ia\sigma}$ annihilating the BCS vacuum, $\gaman_{\ia\sigma}\ket{0}=0$, and it can be diagonalized exactly. After diagonalizing the lead space matrix $\m{M}_l$ we obtain two decoupled channels $\delta=1,2$
\begin{subequations}\label{chandelta}
\begin{equation}
\begin{aligned}
&\gaman_{1\bk\sigma}=a\gaman_{L\bk\sigma}+b\e^{+\i\phi/2}\gaman_{R\bk\sigma},\\
&\gaman_{2\bk\sigma}=a\gaman_{R\bk\sigma}-b\e^{-\i\phi/2}\gaman_{L\bk\sigma},\\
\end{aligned}
\end{equation}
where
\begin{equation}
\begin{aligned}
&a=\sqrt{\frac{1}{2}\left(1+\frac{J_{LL}-J_{RR}}{J_{d}}\right)},\\
&b=\sqrt{\frac{1}{2}\left(1-\frac{J_{LL}-J_{RR}}{J_{d}}\right)}.
\end{aligned}
\end{equation}
The corresponding eigenvalues for two channels are
\begin{equation}\label{thechannels}
\begin{aligned}
&J_{1/2}=\frac{1}{2}\left(J_{LL}+J_{RR}\pm J_{d}\right),\\
&J_{d}=\sqrt{\left(J_{LL}-J_{RR}\right)^2+4\abs{J_{LR}}^2\cos^2\frac{\phi}{2}}.
\end{aligned}
\end{equation}
\end{subequations}
The eigenstates of the spin matrix
\begin{equation}\label{fiwhs}
\m{M}_{s,\delta}=
\bordermatrix{&\ket{\up_{\delta\bk},\upbl}   &\ket{\up_{\delta\bk},\downbl} &\ket{\down_{\delta\bk},\upbl} &\ket{\down_{\delta\bk},\downbl} \cr
                &\frac{1}{2}    &0             &0            &0\cr
                &0            &-\frac{1}{2}     &1 &0   \cr
                &0            &1  &-\frac{1}{2}    &0   \cr
                &0 &0             &0            &\frac{1}{2}   },
\end{equation}
are the singlet
\begin{subequations}\label{STbasis}
\begin{equation}
\ket{S_{\delta\bk}}=\frac{1}{\sqrt{2}}\left(\ket{\up_{\delta\bk},\downbl}-\ket{\down_{\delta\bk},\upbl}\right),
\quad \lambda_{S}=-\frac{3}{2},
\end{equation}
and triplet states
\begin{equation}
\begin{aligned}
&\ket{T^{0}_{\delta\bk}}=\frac{1}{\sqrt{2}}\left(\ket{\up_{\delta\bk},\downbl}+\ket{\down_{\delta\bk},\upbl}\right),\\
&\ket{T^{+}_{\delta\bk}}=\ket{\up_{\delta\bk},\upbl}, \quad \ket{T^{-}_{\delta\bk}}=\ket{\down_{\delta\bk},\downbl},
\quad \lambda_{T}=\frac{1}{2}.
\end{aligned}
\end{equation}
\end{subequations}
Here $\lambda_{S/T}$ denotes the corresponding eigenvalues and
\begin{equation}\label{ceSnot}
\ket{\sigma_{\delta\bk},\etabl}=\gamd_{\delta\bk\sigma}\ket{0}\ket{\etabl},
\text{ with } S^{z}\ket{\etabl}=\etabl/2\ket{\etabl}.
\end{equation}
We note that the states \eqref{STbasis} and \eqref{ceSnot} span only the Hilbert sub-space of single particle excitations with respect to the ground-state doublet $\ket{D_{s}}=\ket{0}\ket{\etabl}$.
Expressed in the singlet/triplet basis~\eqref{STbasis}, the Hamiltonian now takes the following simple form
\begin{equation}\label{shamLRJ}
\begin{aligned}
&H_{\LR}+H_{J}=\sum_{\delta\bk}E_{\bk}\left(\ket{S_{\delta\bk}}\bra{S_{\delta\bk}}+\Sigma_{j}\ket{T^{j}_{\delta\bk}}\bra{T^{j}_{\delta\bk}}\right)\\
&\phantom{...}-\frac{3}{2}\sum_{\delta\bk'\bk}J_{\delta}\ket{S_{\delta\bk'}}\bra{S_{\delta\bk}}
+\frac{1}{2}\sum_{j\delta\bk'\bk}J_{\delta}\ket{T^{j}_{\delta\bk'}}\bra{T^{j}_{\delta\bk}}.
\end{aligned}
\end{equation}

To find a singlet subgap state from the above Hamiltonian \eqref{shamLRJ} we form the linear superposition
\begin{equation}
\ket{S_{\delta}}=\sum_{\bk}A_{\delta\bk}\ket{S_{\delta\bk}},
\end{equation}
and solve the stationary Schr\"{o}dinger equation
\begin{equation}\label{schSeq}
(H_{\LR}+H_{J}-E)\ket{S_{\delta}}=0.
\end{equation}
Projecting Eq.~\eqref{schSeq} to $\bra{S_{\delta\bkap}}$ we obtain the equation
\begin{equation}\label{sceqsc}
A_{\delta\bkap}=\frac{3J_{\delta}}{2}\frac{\sum_{\bk}A_{\delta\bk}}{E_{\bkap}-E},
\end{equation}
which is integrated over $\bkap$ to yield
\begin{equation}
1=3g_{\delta}I_{E},
\end{equation}
where $g_{\delta}=\pi\nu_F J_{\delta}/2$ and the necessary integral, $I_{E}$, for subgap states with $\abs{E}<\absDelta$ and large bandwidth $D\gg\absDelta$ is given by
\begin{equation}
\begin{aligned}
I_{E}&=\frac{1}{\pi\nu_F}\sum_{\bkap}\frac{1}{E_{\bkap}-E}\\
&\approx \frac{2}{\pi}\ln\absB{\frac{2D}{\Delta}}
+\frac{2E\left(\frac{1}{2}+\frac{1}{\pi}\arcsin\frac{E}{\Delta}\right)}{\sqrt{\absDelta^2-E^2}}.
\end{aligned}
\end{equation}
We note that the subgap triplet solutions can exist for ferromagnetic coupling, $g_{\delta}<0$.~\cite{Soda1967,Nagaoka1971,Kirsanskas2014}
However, the Anderson model always gives rise to antiferromagnetic exchange, and therefore there will be no triplet subgap state. Parameterizing the leading-order energy of singlet solutions as
\begin{equation}\label{eade}
E_{0\delta}=\absDelta(1-\eta_{0\delta}^2),
\end{equation}
we finally obtain the perturbative solution valid to lowest (second) order in $g_{\delta}$
\begin{align}\label{etaSTbz}
\abs{\eta_{0\delta}}=
3\sqrt{2}g_{\delta}.
\end{align}
This result matches the second order expansion (in $g_{\alpha\alpha'}$) of the excitation energy obtained from the polarized-spin approximation for $w=0$, $B=0$, and $S=\frac{1}{2}$, when the replacement $g_{\alpha\alpha'}\rightarrow 3g_{\alpha\alpha'}$ is made in \eqref{esosaE}. This could be anticipated already by comparing the expectation value of $H_{J}$ for the state $\ket{\up_{\delta\bk},\downbl}$ corresponding to a polarized spin and for the state $\ket{S_{\delta\bk}}$:
\begin{equation}
\frac{\bra{S_{\delta\bk}}H_{J}\ket{S_{\delta\bk}}}{\bra{\up_{\delta\bk},\downbl}H_{J}\ket{\up_{\delta\bk},\downbl}}=3.
\end{equation}
If the potential scattering term, $H_{W}$, is included, the perturbative result and the polarized-spin approximation still match (see Ref.~\onlinecite{Kirsanskas2014}).

\subsection{\label{sec:qbfield}Magnetic field dependence}

In the singlet/triplet basis, the Zeeman term takes the following form
\begin{equation}
\begin{aligned}
H_{B}=&-\tilde{B}\sum_{\delta\bk}
\left(\ket{S_{\delta\bk}}\bra{T^{0}_{\delta\bk}}+\ket{T^{0}_{\delta\bk}}\bra{S_{\delta\bk}}\right) \\
&+\bar{B}\sum_{\delta\bk}\left(\ket{T^{+}_{\delta\bk}}\bra{T^{+}_{\delta\bk}}
-\ket{T^{-}_{\delta\bk}}\bra{T^{-}_{\delta\bk}}\right),\\
&+\frac{g_{\mri}B}{2}\left(\ket{D_{\upbl}}\bra{D_{\upbl}}-\ket{D_{\downbl}}\bra{D_{\downbl}}\right),
\end{aligned}
\end{equation}
where the last term represents the Zeeman splitting of the ground-state doublet $\ket{D_{\etabl}}=\ket{0}\ket{\etabl}$. In terms of the two potentially different $g$-factors, we have introduced the difference and average $B$-fields as
\begin{equation}
\tilde{B}=\frac{B}{2}(g_{\mri}-g_{\mrce}), \quad
\bar{B}=\frac{B}{2}(g_{\mri}+g_{\mrce}).
\end{equation}
For $g_{d}\neq g_{\mrce}$, we have $\tilde{B}\neq 0$ and the singlet and the triplet are mixed to form a new eigenstate
\begin{equation}
\ket{\psi_{\delta}}=\sum_{\bk}\left(a_{\delta\bk}\ket{S_{\delta\bk}}
+b_{\delta\bk}\ket{T^{0}_{\delta\bk}}\right).
\end{equation}
Projecting the stationary Schr\"{o}dinger equation
\begin{equation}\label{schSeqB}
(H_{\LR}+H_{B}+H_{J}-E)\ket{\psi_{\delta}}=0
\end{equation}
to the sub-space spanned by $\ket{S_{\delta\bkap}}$ and $\ket{T^{0}_{\delta\bkap}}$ and integrating over $\bq$ yields the secular equation
\begin{equation}\label{magSec}
\begin{vmatrix}
1-\frac{3g_{\delta}}{2}[I_{E+\tilde{B}}+I_{E-\tilde{B}}] & -\frac{g_{\delta}}{2}[I_{E+\tilde{B}}-I_{E-\tilde{B}}] \\
\frac{3g_{\delta}}{2}[I_{E+\tilde{B}}-I_{E-\tilde{B}}] & 1+\frac{g_{\delta}}{2}[I_{E+\tilde{B}}+I_{E-\tilde{B}}]
\end{vmatrix}=0.
\end{equation}

We start by examining Eq.~\eqref{magSec} in the low magnetic field limit where $\abs{\tilde{B}}\ll \abs{g_{\delta}^2\Delta}$. Parameterizing the energy as before,
\begin{equation}
E_{\delta}=\absDelta(1-\eta^2_{\delta}),
\end{equation}
we first expand Eq.~\eqref{magSec} to lowest order in
\begin{equation}
\eta_{\pm}=\sqrt{\eta^2_{\delta}\pm\frac{\tilde{B}}{\absDelta}},
\end{equation}
to obtain the equation
\begin{equation}\label{epemeq}
\eta_{+}\eta_{-}-\sqrt{2}g_{\delta}(\eta_{+}+\eta_{-})-6g_{\delta}^2=0.
\end{equation}
Expanding now to lowest order in $\tilde{B}/(\eta^2\absDelta)$, this yields
\begin{equation}\label{foetal}
\eta^4_{\delta}-2\sqrt{2}g_{\delta}\eta^3_{\delta}-6g^2_{\delta}\eta^2_{\delta}-
\left(\frac{\tilde{B}}{2\absDelta}\right)^2=0,
\end{equation}
which has the following leading order perturbative solution
\begin{equation}
\abs{\eta_{\delta}}=3\sqrt{2}g_{\delta}\Big(1+\frac{3}{16}\frac{\tilde{B}^2}{\eta_{0\delta}^4\absDelta^2}\Big).
\end{equation}
This shows that the energy of the subgap state decreases quadratically with $\tilde{B}$ for small magnetic fields.

In the high field limit where $\abs{\tilde{B}}\gg \abs{g_{\delta}^2\Delta}$, we find from Eq.~\eqref{magSec} to lowest order in $g_{\delta}$ that
\begin{equation}\label{hmgfl}
E_{\delta}=\absDelta(1-\eta_{\delta,\mathrm{cl}}^2)-\tilde{B}, \quad \abs{\eta_{\delta,\mathrm{cl}}}\approx\sqrt{2}g_{\delta},
\end{equation}
which corresponds to neglecting off-diagonal terms in the spin matrix \eqref{fiwhs}. For intermediate magnetic field strengths, the perturbative (in $g_{\delta}$) solution is obtained by numerically solving Eq.~\eqref{epemeq}.

\begin{figure}
\begin{center}
\includegraphics[width=0.8\columnwidth]{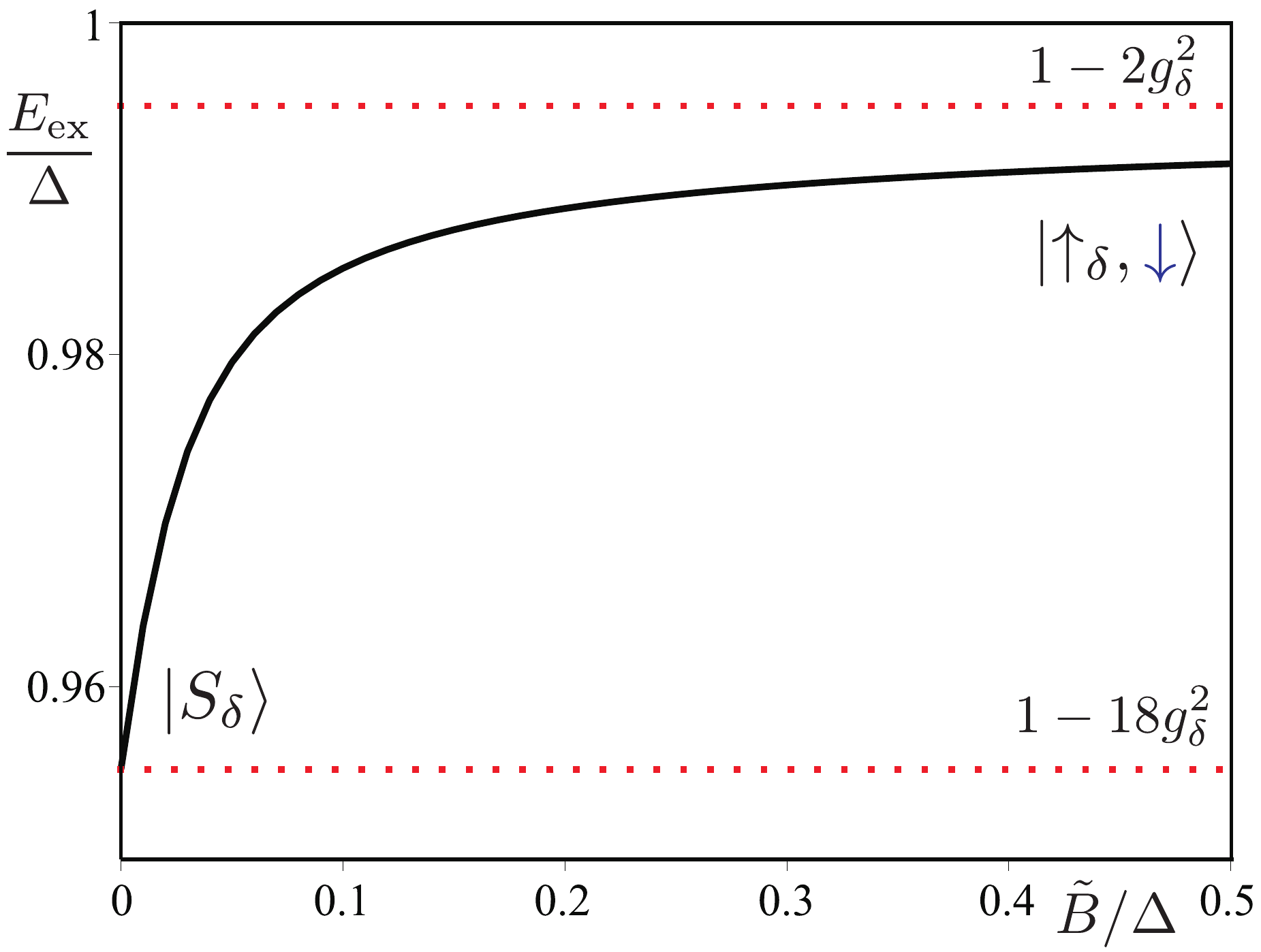}
\caption{\label{fig7}(Color online) Subgap excitation spectrum, $E_{\mathrm{ex}}=E_{\delta}-E_{\downbl}$, with respect to the ground-state doublet $\ket{D_{\downbl}}$ for particular channel $\delta=1,2$. It is assumed that the magnetic field in the superconductor is screened $g_{\mathrm{ce}}\approx 0$. For antiferromagnetic coupling, $g_{\delta}=0.05$, the excited state is singlet like, which for high magnetic fields becomes polarized-spin like $\ket{\up_{\delta},\down}$.}
\end{center}
\end{figure}

Finally, we examine the subgap excitation spectrum in the case where the magnetic field in the superconducting leads is screened, i.e., for $g_{\mrce}\approx 0$ and $\tilde{B}=\bar{B}=g_{\mri}B/2$. Note that we  assume the magnetic field to be much weaker than the critical field and hence neglect its influence on the gap. For positive magnetic field, $\tilde{B}>0$, the ground state is the lower energy component of the doublet with $E_{\downbl}=-\tilde{B}$ and the resulting subgap excitation energy, $E_{\mathrm{ex}}=E_{\delta}-E_{\downbl}$, for a particular channel $\delta$ is depicted in \figurename~\ref{fig7}. For high magnetic fields, this excitation approaches the energy of the polarized spin approximation, Eq.~\eqref{esosaE}. In this case the eigenstate is $\ket{\up_{\delta},\downbl}=\sum_{\bk}A_{\delta\bk}^{\up}\ket{\up_{\delta\bk},\downbl}$, where $A_{\delta\bk}^{\up}$ is determined by projecting the Sch\"{o}dinger equation to $\ket{\up_{\delta\bk},\downbl}$ with neglected spin-flip terms $S^{+}$ and $S^{-}$ in the exchange Hamiltonian (\ref{shamJ}). Changing the sign of the magnetic field simply reverses all spins in the previous discussion.

\subsection{\label{sec:qbeyond}Beyond the single-quasiparticle approximation}

So far we have examined the excitation spectrum when only a single quasiparticle is included and the system is effectively described by the Hamiltonian \eqref{shamJ}. We now return to the original Hamiltonian and investigate the effects of the terms $\gamd_{a'\up}\gamd_{a\down}$ which were neglected to arrive at \eqref{shamJ}. With these terms included, second order perturbation theory yields the following shift of the ground-state energy from which the supercurrent can be deduced
\begin{equation}\label{soecfgd}
\begin{aligned}
&E^{(2)}_{\etabl}=\sum_{l\neq D_{\etabl}}\frac{\abs{\bra{l}H'\ket{D_{\etabl}}}^2}{E_{\etabl}^{(0)}-E_{l}}\\
&=-\sum_{\substack{\alpha k \\ \alpha' k'}}\Big(
\frac{1}{4}\frac{J_{\alpha'\alpha}J_{\alpha\alpha'}\abs{u_{\alpha'k'}v_{\alpha k}\e^{\i\phi_{\alpha}}-u_{\alpha k}v_{\alpha'k'}\e^{\i\phi_{\alpha'}}}^2}{E_{\alpha k}+E_{\alpha'k'}}\\
&\phantom{............}
+\frac{1}{2}\frac{J_{\alpha'\alpha}J_{\alpha\alpha'}\abs{u_{\alpha'k'}v_{\alpha k}\e^{\i\phi_{\alpha}}-u_{\alpha k}v_{\alpha'k'}\e^{\i\phi_{\alpha'}}}^2}{E_{\alpha k}+E_{\alpha'k'}-\etabl(g-g_{ce})B}\\
&\phantom{............}
+\frac{W_{\alpha'\alpha}W_{\alpha\alpha'}\abs{u_{\alpha'k'}v_{\alpha k}\e^{\i\phi_{\alpha}}+u_{\alpha k}v_{\alpha'k'}\e^{\i\phi_{\alpha'}}}^2}{E_{\alpha k}+E_{\alpha'k'}}\Big),
\end{aligned}
\end{equation}
with $\ket{l}$ denoting all possible intermediate states and $H'=H_{J}+H_{W}$.

This second order shift in ground-state energy would appear to influence the observable excitation energies, but when analyzed in terms of a three quasiparticle Yosida wavefunction ansatz (see Ref.~\onlinecite{Kirsanskas2014}), the eigenenergies of the subgap states are found to be shifted in exactly the same way, and overall the shift~\eqref{soecfgd} drops out and the previously obtained second order result for the excitation energies, Eq.~\eqref{etaSTbz}, remains valid. In this manner, the Yosida wave-function ansatz generates a well defined perturbative expansion for the energy differences in the dimensionless couplings, $g$ and $w$.

\subsection{\label{sec:qscrt}Supercurrent}

Going beyond the polarized-spin approximation, we calculate the supercurrent perturbatively from Eqs.~\eqref{degsdphi} and \eqref{soecfgd}. For $B>0$ the ground state is $\ket{D_{\downbl}}$ and one finds
\begin{equation}\label{Isftqs}
I_{S}=2\absDelta\sin^2(2\theta)\left[\left(w^2-g^2\right)F(0)-2g^2F(\tilde{B})\right]\sin\phi,
\end{equation}
to leading order in $w$ and $g$ and with
\begin{equation}
F(\tilde{B})=\frac{1}{\absDelta\pi^2\nu_F^2}\sum_{\bk\bk'}
\frac{u_{\bk}v_{\bk}u_{\bk'}v_{\bk'}}{E_{\bk}+E_{\bk'}+\tilde{B}}
\end{equation}
whereby $F(0)\approx 1/4$ for $D\gg\absDelta$. For zero magnetic field, the supercurrent simplifies to
\begin{align}\label{ISquant}
I_{S}=\frac{\absDelta}{2}\sin^2(2\theta)\left(w^2-3g^2\right)\sin\phi,
\end{align}
which always corresponds to a $\pi$-junction since $g>w$.

For small magnetic fields with $\abs{\tilde{B}}\ll\absDelta$, we have
\begin{align}
F(\tilde{B})-F(0)\approx -\frac{\tilde{B}}{2\pi^2\absDelta}.
\end{align}
From Eq.~\eqref{Isftqs}, it then follows that a positive $\tilde{B}$ decreases, and a negative $\tilde{B}$ increases the magnitude of the supercurrent, $\abs{I_S}$. Notice that $\tilde{B}$ can become negative for positive $B$ when $g_{\mrce}>g_{\mri}$.

For large magnetic fields with $|\tilde{B}|\gg\Delta$, one finds instead
\begin{align}
F(\tilde{B})\approx \frac{\Delta}{\tilde{B}}\left(\ln^{2}\left(\frac{2\tilde{B}}{\Delta}\right)+\frac{\pi^{2}}{6}\right),
\end{align}
and as $F(\tilde{B})$ vanishes with increasing field, the spin is polarized, and the supercurrent~\eqref{ISclas}, obtained within the polarized-spin approximation, is recovered.


%

\end{document}